\documentclass{elsart}
\usepackage{graphicx,amssymb}
\journal{Physic Letters B}
\newcommand\beq{\begin{eqnarray}}
\newcommand\eeq{\end{eqnarray}}
\newcommand\bq{\begin{equation}}
\newcommand\eq{\end{equation}}

\begin{document}

\begin{frontmatter}

\title{Decay of polarized muon at rest as a source of polarized neutrino beam}

\author[UW]{S. Ciechanowicz},
 \ead{ciechano@sunflower.ift.uni.wroc.pl}
\author[UW]{W. Sobk\'ow\corauthref{cor}},
\corauth[cor]{Corresponding author}
 \ead{sobkow@sunflower.ift.uni.wroc.pl}
\author[UJ]{M. Misiaszek}
 \ead{misiaszek@zefir.if.uj.edu.pl}
\address[UW]{Institute of
Theoretical Physics, University of Wroc\l{}aw, Pl. M. Born 9,
\\PL-50-204~Wroc\l{}aw, Poland}
\address[UJ]{ M. Smoluchowski
Institute of Physics, Jagiellonian University, ul. Reymonta 4,\\
PL-30-059 Krak\'ow, Poland}
\begin{abstract}

In this paper, we indicate the theoretical possibility of using
the decay of polarized muons at rest as a source of the
transversely polarized electron antineutrino beam. Such a beam can
be used to probe new effects beyond standard model. We mean here
new tests concerning  CP violation, Lorentz structure and
chirality structure of the charged current weak interactions. The
main goal is to show how the energy and angular distribution of the
electron antineutrinos in  the muon rest frame depends on the transverse components of the  antineutrino beam polarization. We admit the participation of the complex exotic  scalar and tensor couplings of the right-chirality
electron antineutrinos in addition to the standard  vector coupling
of  the left-chirality ones, while the muon neutrinos are always
left-chirality. It means that the outgoing electron antineutrino
beam is a mixture of the left- and right-chirality antineutrinos and
has the fixed direction of the transverse spin polarization with
respect to the production plane.
 Our analysis is model-independent
and consistent with the current upper limits on the non-standard
couplings. The results are presented in a limit of infinitesimally
small mass for all particles produced in the decay.

\end{abstract}

\begin{keyword}
polarized muon decay \sep exotic couplings \sep transverse antineutrino spin polarization 
\PACS 13.15.+g \sep 13.35.Bv \sep 13.88.+e 
\end{keyword}
\end{frontmatter}
\section{Introduction}
\label{sec1}

 Decay of polarized muon at rest (DPMaR) is the very appropriate process to test
both the time reversal violation (TRV) and the space-time, and
chirality structure of the purely leptonic charged weak interactions
(PLCWI). Moreover, we investigate here if the DPMaR can also be used
as a strong source of the transversely polarized antineutrino
(neutrino) beam which would be scattered on the polarized electron
target (PET). A detailed analysis of new effects beyond the Standard
Model (SM) of electroweak interactions \cite{Glashow,Wein,Salam} is
carried out in \cite{azpet}. Ciechanowicz et al. show that  the
scattering of the left-chirality and longitudinally polarized
neutrino  beam on the PET would allow to test the possibility of
CP-breaking  in the $(\nu_\mu e^{-})$ scattering. The measurement of
the azimuthal asymmetry of recoil electrons could detect the
CP-violating phase between the standard complex vector and
axial-vector couplings. The other problem considered in \cite{azpet}
concerns the scattering of the transversely polarized mixture of
left-chirality and right-chirality neutrinos  on the PET. If such a
neutrino beam would be scattered, the dependence on the angle
between the transverse neutrino spin polarization of incoming beam
and the transverse electron polarization of target in the recoil
electron energy spectrum could be tested. That would be a direct
signature of the right-chirality neutrinos.
 \\ As is well-known, the SM of electro-weak
interactions  has a vector-axial (V-A) structure \cite{Gell} which
has been put by hand in order to obtain  agreement with experiments.
This means that only left-chirality Dirac neutrinos may take part in
the charged and neutral current weak interaction. This structure
follows among other from the low energy measurements of the muon
decay such as; the spectral shape, angular distribution, and
polarization of the outgoing electrons (positrons). At present,
there is no evidence for the deviations from the SM for the Michel
parameters.  The neutrino oscillation experiments indicate the
non-zero neutrino mass and provide first evidence for physics beyond
the minimal SM. On the other hand, the experimental precision of
present tests still allows the participation of the exotic scalar S,
tensor T and pseudoscalar P couplings of the right-chirality Dirac
neutrinos beyond the  SM \cite{Delphi}. The KARMEN experiment
\cite{KARMEN} has measured the energy distribution of  electron
neutrinos emitted in positive muon decay at rest $(\mu^+ \rightarrow
e^+ + \nu_e + \overline{\nu}_\mu)$. The obtained result is in
agreement with the SM prediction on the neutrino Michel parameter
$\omega_L=0$. They get for the first time  a $90\%$ confidence upper
limit of $\omega_L\leq 0.113$, which leads to a limit of $|g_{RL}^S
+ 2g_{RL}^T|\leq 0.78$ for the interference between the scalar and
tensor couplings. The current upper limits on the all non-standard
couplings, obtained from the normal and inverse muon decay, are
presented in the Table \ref{table1} \cite{Data}. The coupling
constants are denoted as $g_{\epsilon \mu}^{\gamma}$, where $\gamma=
S, V, T$ indicates the type of weak interaction, i.e. scalar S,
vector V, tensor T; $\epsilon, \mu=L, R$ indicate the chirality of
the electron or muon and the neutrino chiralities
 are uniquely determined for given $\gamma, \epsilon, \mu$. It
means that the neutrino chirality is the same as the associated
charged lepton for the V interaction, and opposite for the S, T
interactions \cite{Data}. In the SM, only $g_{LL}^{V}$ is non-zero value.
\begin{table}
\begin{center}
%
\begin{tabular}{|c|c|c|}
  \hline  
  Coupling constants & SM & Current limits \\
  \hline
   $|g_{LL}^V|$ & $1$ &   $>0.960$ \\
    $ |g_{LR}^V|$ & $0$ & $<0.036$ \\
    $|g_{RL}^V|$ & $0$ & $<0.104$ \\
    $|g_{RR}^V|$ & $0$ & $<0.034$ \\
    \hline
  $|g_{LL}^{S}|$ & 0 & $<0.550$\\
  $|g_{LR}^{S}|$& 0 & $<0.088$\\
  $|g_{RL}^{S}|$& 0 & $<0.417$\\
  $|g_{RR}^{S}|$& 0 & $<0.067$\\
  \hline
   $|g_{LL}^{T}|$& 0 & $0$\\
   $|g_{LR}^{T}|$& 0 & $<0.025$\\
   $|g_{RL}^{T}|$& 0 & $<0.104$\\
   $|g_{RR}^{T}|$& 0 & $0$\\
  \hline 
\end{tabular}
\caption{\label{table1} Current limits on the non-standard
couplings.}
%
\end{center}
\end{table}

It is necessary to point out  that  the existence of the exotic
right-chirality neutrinos in the few keV region, that are sterile in
SM, can have numerous consequences in  astrophysics and cosmology.
We mean here the mechanism of neutrino {}``spin flip'' in the Sun's
convection zone in order to explain the observed deficit  of the
solar neutrinos \cite{key-1}. In addition, the sterile neutrinos
could also account for pulsar kicks (high pulsar velocities), could
explain all or some fraction of the dark matter in the Universe and
would affect emission of supernova  neutrinos \cite{Biermann}.
\\ Recent analysis carried out by Erwin et al. for the muon decay \cite{Erwin}
shows that there exist
 four-fermion operators that do not contribute  to the neutrino mass
 matrix through  radiative corrections. These operators generate the
 exotic couplings $g_{LR,RL}^{S,T}$, while all operators generating
 the  vector couplings $ g_{LR,RL}^{V}$ contribute to the neutrino
 mass matrix.
\\ So far the CP violation is observed only in the decays of
neutral K- and B-mesons \cite{CP}, and is described by a single
phase of the Cabibbo-Kobayashi-Maskawa quark-mixing matrix
(CKM)\cite{Kobayashi}. There is no experimental evidence on the TRV
in the PLCWI, e.g. the muon decay and neutrino-electron elastic
scattering. However, the baryon asymmetry of the Universe can not be
explained by the CKM phase only, and new sources of the CP violation
are required \cite{barion}. According to the prediction of
non-standard models, the effects of new CP-breaking phases could be
measured in  observables where the SM CP-violation is suppressed,
while alternative sources can generate a sizable effect, e.g. the
electric dipole moment of the neutron, the transverse lepton
polarization in three-body decays of charged kaons $K^+$ \cite{Lee,
Weinb}, transverse polarization of the electrons emitted in the
decay of polarized $^8Li$ nuclei \cite{Huber}.
\\ The other possibility of measuring the exotic CP-breaking
phases is to use the neutrino observables which consist only of the
interference terms between the standard coupling of the
left-chirality neutrinos and exotic couplings  of the
right-chirality neutrinos and do not depend on the neutrino mass. We
mean here both T-even and T-odd transverse components of the
neutrino spin polarization. At present, the direct tests are still
impossible. The possible solution can be the scattering of the
transversely polarized (anti)neutrino beam, coming from the
polarized muon decay at rest, on the PET and the measurement of the
maximal asymmetry of the cross section, \cite{azpet}. \\
Left-right symmetric models (LRSM) and  composite models (CM) can be
proposed  as an example of the non-standard models of purely
leptonic weak interactions, in which the exotic couplings of the
right-chirality neutrinos (antineutrinos) can appear.
Recently TWIST Collaboration \cite{Twist} has measured the
 Michel parameter $\rho$ in the normal $\mu^+$ decay and has set new
 limit on the $W_L - W_R$ mixing angle in the LRSM. Their result $\rho= 0.75080
 \pm 0.00044 (stat.) \pm 0.00093 (syst.) \pm 0.00023$ is in good
 agreement with the SM prediction $\rho=3/4$,  and sets new upper limit
 on mixing angle $|\chi| < 0.030 \; (90 \% \; CL)$. The
CM have been proposed to probe the scale for compositeness of quarks
and leptons.   Lagrangian of the new effective contact interactions
(CI) for the muon decay includes among other the contributions from
the standard vector coupling of the left-chirality neutrinos and
exotic scalar coupling of the right-chirality ones \cite{CM}.
\\ Our analysis is model-independent and  the calculations are made
in the limit of infinitesimally small mass for all particles
produced in the muon  decay. The density operators  \cite{Michel}
for the polarized initial muon and for the polarized outgoing
electron antineutrino are used, see Appendix A. We use the system of natural units
with $\hbar=c=1$, Dirac-Pauli representation of the
$\gamma$-matrices and the $(+, -, -, -)$ metric \cite{Greiner}.

\begin{figure}
\label{pmu}
\begin{center}
\includegraphics*[scale=.5]{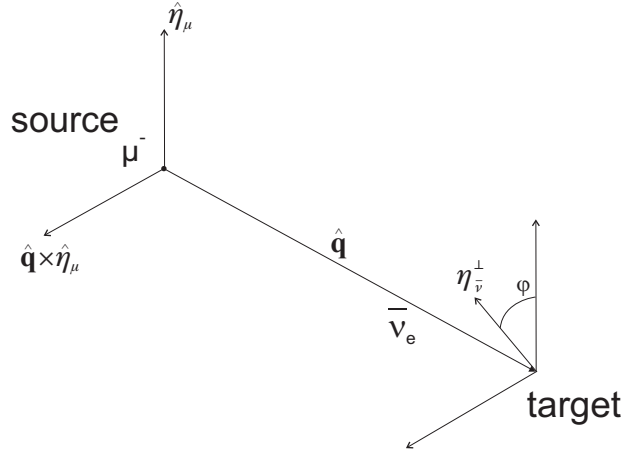}
\end{center}
\caption{Figure shows the production  plane of the
$\overline{\nu}_{e}$-antineutrinos for the process of $(\mu^{-}
\rightarrow e^- + \overline{\nu}_{e} + \nu_{\mu})$, $\mbox{\boldmath
$\eta_{\overline{\nu}}^\perp$}$ - the transverse  polarization of the outgoing
antineutrino. The production plane is spanned by the direction of the
initial muon polarization $\mbox{\boldmath $\hat{\eta}_{\mu}$}$ and
of the outgoing antineutrino momentum ${\bf \hat{ q}}$.}

\end{figure}

\begin{table}
\begin{center}
\begin{tabular}{|c|c|c|c|}
 \hline 
SM  & Nonstandard   & Interference of & Interference of \\
 & couplings (NC)  & SM with NC & SM with NC \\
\cline{1-4}  & & a & b \\
\cline{1-4}
$g_{LL}^{V}$ & $g_{LL}^{S}$ & $g_{LL}^{V}g_{LL}^{S*} = 0$ & $g_{LL}^{V}g_{LL}^{S*} = 0$  \\
& $g_{LR}^{S}$ & $g_{LL}^{V}g_{LR}^{S*}\not = 0 $ & $g_{LL}^{V}g_{LR}^{S*} = 0 $\\
& $g_{RL}^{S}$ & $g_{LL}^{V}g_{RL}^{S*}=0$ & $g_{LL}^{V}g_{RL}^{S*} = 0$\\
& $g_{RR}^{S}$ & $g_{LL}^{V}g_{RR}^{S*}=0$ & $g_{LL}^{V}g_{RR}^{S*} = 0$\\
& $g_{LL}^{T}$ & $g_{LL}^{V}g_{LL}^{T*} =0 $ & $g_{LL}^{V}g_{LL}^{T*} = 0$\\
& $g_{LR}^{T}$ & $g_{LL}^{V}g_{LR}^{T*} \not=0$ & $g_{LL}^{V}g_{LR}^{T*} = 0$\\
& $g_{RL}^{T}$ & $g_{LL}^{V}g_{RL}^{T*} =0$ & $g_{LL}^{V}g_{RL}^{T*} = 0$\\
& $g_{RR}^{T}$ & $g_{LL}^{V}g_{RR}^{T*} =0$ & $g_{LL}^{V}g_{RR}^{T*} = 0$\\
 \hline 
\end{tabular}
\caption{Table shows non-vanishing  interferences
between the $g_{LL}^{V}$ and $g_{\epsilon \mu}^{S,T} \; (\epsilon
\mu= LL, LR, RL, RR)$ couplings  after a computation of traces for
two cases: a. Polarized muon and polarized electron antineutrino. b.
Polarized muon and polarized muon neutrino.}
\label{table2}
\end{center}
\end{table}

\section{ Transversely Polarized Electron-Antineutrino Beam}
\label{Sec2}

To show how the  energy and angular distribution of the
electron antineutrinos may depend on the angle
between the transverse antineutrino spin polarization and the
muon polarization vectors, we assume that the DPMaR  $(\mu^{-}
\rightarrow e^- + \overline{\nu}_{e} + \nu_{\mu})$ is a source of
the electron antineutrino beam. It is worth to notice that if one
takes into account the positive muon decay $(\mu^{+} \rightarrow e^+
+ \overline{\nu}_{\mu} + \nu_{e})$, the electron neutrino beam is 
produced. The production plane is spanned by the direction of the
initial muon polarization $\mbox{\boldmath $\hat{\eta}_{\mu}$}$ and
of the outgoing electron antineutrino momentum ${\bf \hat{q}} $,
Fig. \ref{pmu}.  We admit a presence  of the exotic scalar
$g_{LR}^S$ and tensor $g_{LR}^T$  couplings in addition to the
standard vector  $g_{LL}^V$ coupling. It means that the outgoing
electron antineutrino flux is a mixture of the left-chirality
antineutrinos produced in the  $g_{LL}^V$ weak interaction and the
right-chirality ones produced in the  $g_{LR}^S$ and the $g_{LR}^T$
weak interactions. Because our analysis is carried out in the limit of vanishing antineutrino mass, the left-chirality antineutrino has positive helicity, while the right-chirality one has negative helicity, see \cite{Fetscher}. The muon neutrino is always left-chirality both
for the $g_{LL}^V$  and the $g_{LR}^S$, $g_{LR}^T$ couplings (muon neutrino has negative helicity, when $m_{\nu_\mu}\rightarrow 0)$. In the
SM, only $g_{LL}^{V}$ is non-zero value. The table \ref{table2}
displays explicitly that there are only two non-zero interferences
between the standard coupling $g_{LL}^V$ and exotic  couplings, i.
e. $g_{LR}^S$ and $g_{LR}^T$. Because we allow for the
non-conservation of the combined symmetry
 CP,  all the  coupling constants $g_{LL}^V, g_{LR}^S, g_{LR}^T $ are complex.
 The  amplitude for the polarized muon decay is of the form:
  \beq M_{\mu^{-}} & = &
\frac{G_{F}}{\sqrt{2}}\{g_{LL}^{V}(\overline{u}_{e}\gamma_{\alpha}(1-\gamma_5)v_{\nu_{e}})
(\overline{u}_{\nu_{\mu}} \gamma^{\alpha}(1 -
\gamma_{5})u_{\mu}) \nonumber\\
&&  \mbox{} + g_{LR}^{S}(\overline{u}_{e} (1+\gamma_5)v_{\nu_{e}})
  (\overline{u}_{\nu_{\mu}}(1 + \gamma_{5})u_{\mu})\\
&&  \mbox{} + \frac{g_{LR}^{T}}{2}(\overline{u}_{e} \sigma_{\alpha
\beta}(1+\gamma_5)v_{\nu_{e}})
  (\overline{u}_{\nu_{\mu}}\sigma^{\alpha
\beta}(1 + \gamma_{5})u_{\mu})\},\nonumber
  \eeq
where $ v_{\nu_{e}}$ and $\overline{u}_{e}$ $(u_{\mu}\;$ and $\;
\overline{u}_{\nu_{\mu}})$ are the Dirac bispinors of the outgoing
electron antineutrino and electron (initial muon and final muon
neutrino), respectively. $G_{F}= 1.16639(1)\times
10^{-5}\,\mbox{GeV}^{-2}$ \cite{Data} is the Fermi constant.
 The coupling constants are
denoted as $g^{V}_{LL}$  and $g^{S}_{LR}, g_{LR}^T$ respectively to
the chirality of the final electron and initial muon.
 The formula for the the energy and angular distribution of the
electron antineutrinos in  the muon rest frame,
including interference terms between the standard $g_{LL}^{V}$ and
exotic $g_{LR}^S, g_{LR}^T$ couplings with $\mbox{\boldmath
$\hat{\eta}_{\mu}$} \cdot {\bf \hat{ q}}\not = 0$ is of the form:
   \beq
   \label{didera}
 \frac{d^2 \Gamma}{ dy d\Omega_\nu}  
 & = & \left(\frac{d^2 \Gamma}{ dy
d\Omega_\nu}\right)_{(V)} + \left(\frac{d^2 \Gamma}{dy
d\Omega_\nu}\right)_{(S + T)} + \left(\frac{d^2 \Gamma}{dy
d\Omega_\nu}\right)_{(VS + VT)}, \\
\left(\frac{d^2 \Gamma}{ dy d\Omega_\nu}\right)_{(V)} & = & \frac{
G_{F}^2 m_{\mu}^5 }{128\pi^4}\Bigg\{|g_{LL}^{V}|^2 y^2 (1 - y)
(1+\mbox{\boldmath $\hat{\eta}_{\overline{\nu}}$}\cdot\hat{\bf
q})(1+\mbox{\boldmath $\hat{\eta}_{\mu}$}\cdot\hat{\bf
q})\Bigg\}, \label{livetimeV}\\
  \left(\frac{d^2 \Gamma}{ dy d\Omega_\nu}\right)_{(S + T)} & = &
\frac{G_{F}^2 m_{\mu}^5  }{3072\pi^4}(1-\mbox{\boldmath
$\hat{\eta}_{\overline{\nu}}$}\cdot\hat{\bf q})|g_{LR}^{S}|^2 y^2\Bigg\{
\bigg[(3 - 2y) - (1 - 2y)\mbox{\boldmath
$\hat{\eta}_{\mu}$}\cdot\hat{\bf q}\bigg] \nonumber\\
&& \mbox{} + 4 \left|\frac{g_{LR}^{T}}{g_{LR}^{S}}\right|^2
\bigg[(15 - 14y) - (13 - 14y)\mbox{\boldmath
$\hat{\eta}_{\mu}$}\cdot\hat{\bf q} \bigg]\Bigg\}, \label{livetimeST} \eeq
\noindent
\beq  
\lefteqn{ \left(\frac{d^2 \Gamma}{dy
d\Omega_\nu}\right)_{(VS + VT)} =  
\frac{G_{F}^2 m_{\mu}^5}{256\pi^4} y^2(1 - y) } \\ 
&& \mbox{} \cdot \Bigg\{  Re(g_{LL}^V g_{LR}^{S*}) (\mbox{\boldmath
$\eta_{\overline{\nu}}^{\perp}$}\cdot \mbox{\boldmath $\hat{\eta}_{\mu}$})  
+ Im(g_{LL}^V g_{LR}^{S*})  \mbox{\boldmath
$\eta_{\overline{\nu}}^{\perp}$}\cdot({\bf \hat{q}} \times \mbox{\boldmath
$\hat{\eta}_{\mu}$}) \nonumber \\ 
&& \mbox{} - 6 \bigg[ Re(g_{LL}^V g_{LR}^{T*}) (\mbox{\boldmath
$\eta_{\overline{\nu}}^{\perp}$}\cdot \mbox{\boldmath
$\hat{\eta}_{\mu}$})  
 + Im(g_{LL}^V g_{LR}^{T*})\mbox{\boldmath
$\eta_{\overline{\nu}}^{\perp}$}\cdot({\bf \hat{q}} \times \mbox{\boldmath
$\hat{\eta}_{\mu}$}) \bigg]\Bigg\},
\label{livetimeVST} \nonumber \eeq
where $m_\mu$ is the muon mass,  $y=\frac{2E_\nu}{m_\mu}$ is the
reduced antineutrino energy, it varies from $0 $ to $1$,
$\mbox{\boldmath $\hat{\eta}_{\overline{\nu}}$}\cdot\hat{\bf q} = + 1 $ is the
longitudinal polarization of the left-chirality electron
antineutrino for the standard $g^{V}_{LL}$, while $\mbox{\boldmath
$\hat{\eta}_{\overline{\nu}}$}\cdot\hat{\bf q} = - 1 $ is the longitudinal
polarization of the right-chirality electron antineutrino for the
exotic $g^{S, T}_{LR}$ couplings. \\  It is necessary to point out
that the above formula is presented after the integration over all
the momentum directions of the outgoing electron and muon neutrino.
If the $\mbox{\boldmath
$\hat{\eta}_{\mu}$} \cdot {\bf \hat{ q}}= 0$ the interference part can be rewritten in the following way:
 \beq
   \label{DDR}
 \left(\frac{d^2 \Gamma}{dy d\Omega_\nu}\right)_{(VS + VT)}
& = & \frac{G_{F}^2 m_{\mu}^5 }{256\pi^4}
   |\mbox{\boldmath $\eta_{\overline{\nu}}^{\perp}$}|
|\mbox{\boldmath $\eta_{\mu} ^{\perp}$}| |g_{LL}^V ||g_{LR}^{S}|\\
&& \mbox{} \cdot \bigg\{  cos(\phi - \alpha_{VS})
-6\left|\frac{g_{LR}^{T}}{g_{LR}^{S}}\right|cos(\phi -
\alpha_{VT})\bigg\}y^2(1 - y), \nonumber \eeq
where $\phi$ is the angle between the direction of $\mbox{\boldmath
$\eta_{\overline{\nu}}^{\perp}$}$ and the direction of $\mbox{\boldmath
$\eta_{\mu} ^{\perp}$}$;   $\alpha_{V S} \equiv \alpha_{V}^{LL} -
\alpha_{S}^{LR}$, $\alpha_{V T} \equiv \alpha_{V}^{LL} -
\alpha_{T}^{LR} $ are the relative phases between the $g^{V}_{LL}$
and $g^{S}_{LR}, g^{T}_{LR}$ couplings. \par
\begin{figure}
\label{SMVST}
\begin{center}
\includegraphics*[scale=0.55,angle=-90]{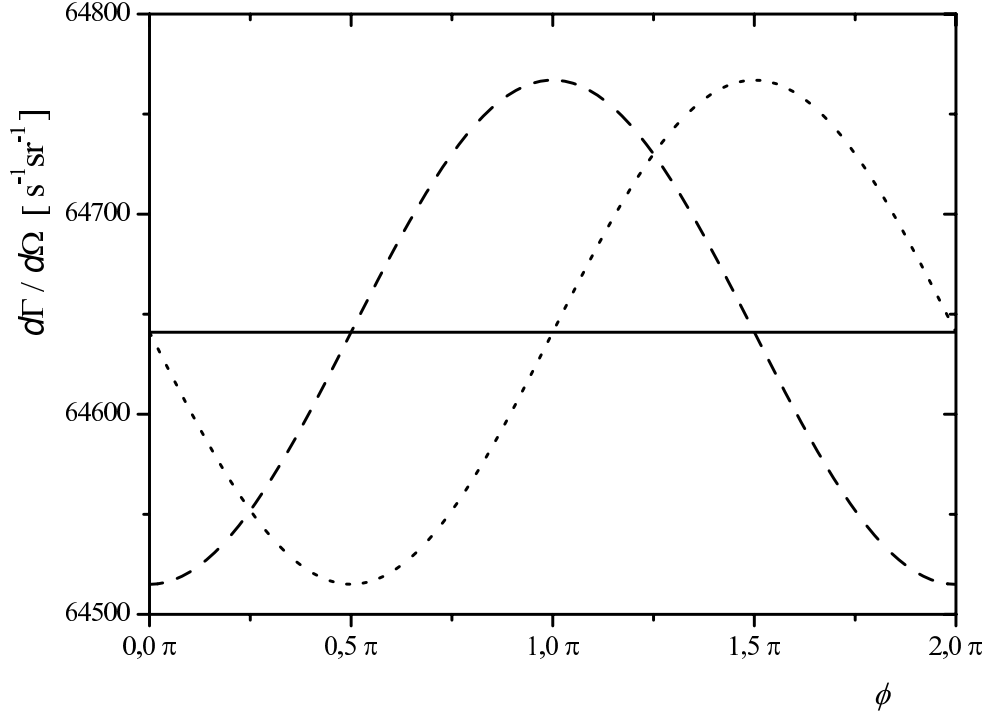}
\end{center}
\caption{Plot of the $\frac{d\Gamma}{ d\Omega_\nu}$
as a function of $\phi$:  
a) solid line is for the SM;  
b) CP violation, 
$\alpha_{VS}=\pi/2, \alpha_{VT}=\pi/2 $  (short-dashed line); 
c) CP conservation,   
$\alpha_{VS}=0, \alpha_{VT}=0$ (long-dashed line)}. 
\end{figure}

 We see that in the case
of the transversely polarized antineutrino beam coming from the 
polarized muon decay, the interference terms between the standard 
coupling $g_{LL}^{V} $ and exotic $g_{LR}^{S, T}$ couplings  do  
not vanish in the limit of vanishing electron-antineutrino and 
muon-neutrino masses. This independence on the
mass makes the measurement of the relative phases $\alpha_{VS},
\alpha_{VT}$ between these couplings possible. The interference
part, Eq. (\ref{DDR}), includes only the contributions from the
transverse component of the initial muon polarization
$\mbox{\boldmath $\eta_{\mu} ^{\perp}$} $ and the transverse
component of the outgoing antineutrino polarization $\mbox{\boldmath
$\eta_{\overline{\nu}}^{\perp}$}$. Both transverse components are perpendicular
with respect to the $\hat{\bf q}$. It can be noticed that the 
relative phases $\alpha_{VS}, \alpha_{VT}$ different from $0, \pi$
would indicate the CP violation in the CC weak interaction. 
 Using the current data \cite{Data}, we calculate the upper limit on the magnitude of the transverse antineutrino polarization and lower bound for the longitudinal antineutrino polarization, see \cite{Fetscher}: 
 \beq 
 \label{trlo}
 |\mbox{\boldmath $\eta_{\overline{\nu} }^{\perp}$}| &=& 2\sqrt{Q_{L}^{\overline{\nu}}(1-Q_{L}^{\overline{\nu}}) } \leq 0.126, \;
 \mbox{\boldmath
$\hat{\eta}_{\overline{\nu}}$}\cdot\hat{\bf q} = 2 Q_{L}^{\overline{\nu}} -1 \geq 0.992,\\
Q_{L}^{\overline{\nu}} &=& 1 - \frac{1}{4}|g_{LR}^S|^{2} - 3 |g_{LR}^T|^{2}\geq 0.996,  
 \eeq
 where $Q_{L}^{\overline{\nu}}$ is the probability of obtaining the
left-chirality (anti)neutrino.\\ 
The Fig. \ref{SMVST} shows the plot of the $\frac{d\Gamma}{ d\Omega_\nu}$
as a function of the azimuthal angle $\phi$ 
for  $ \mbox{\boldmath $\hat{\eta}_{\mu}$}\cdot\hat{\bf q}=0, \mbox{\boldmath
$\hat{\eta}_{\overline{\nu}}$}\cdot\hat{\bf q}=0.992, |\mbox{\boldmath $\eta_{\overline{\nu} }^{\perp}$}|=0.126, |\mbox{\boldmath $\eta_{\mu} ^{\perp}$}|=1, y=2/3, |g_{LR}^S|=0.088, |g_{LR}^T|=0.025, |g_{LL}^V|=0.998$. The short-dashed line illustrates  the possible effect of the CP violation for the relatives phases $\alpha_{VS}=\pi/2, \alpha_{VT}=\pi/2$, while the long-dashed line  represents the case of the CP conservation  for the  $\alpha_{VS}=0, \alpha_{VT}=0$.  \\
We note that the Eq. (\ref{livetimeV}) after integration over all the
antineutrino directions (with $|g_{LL}^V |=1, \mbox{\boldmath
$\hat{\eta}_{\overline{\nu}}$}\cdot\hat{\bf q} = +1$) is the same as the Eq.
(7) in \cite{Fetscher}  (with  $Q^\nu_L =1, \omega_L=0, \eta_L=0$,
neglecting the masses of the neutrinos and of the electron as well as radiative corrections). 
 We see that for $\mbox{\boldmath 
$\hat{\eta}_{\mu}$}\cdot\hat{\bf q}=-1$ only  the exotic part $(S + T)$ survives:
\beq \left(\frac{d^2 \Gamma}{ dy
d\Omega_\nu}\right)_{(S + T)} & = & \frac{G_{F}^2 m_{\mu}^5
}{768\pi^4}(1-\mbox{\boldmath $\hat{\eta}_{\overline{\nu}}$}\cdot\hat{\bf
q})|g_{LR}^{S}|^2 y^2(1 - y) \Bigg\{ 1 + 28 \left|
\frac{g_{LR}^{T}}{g_{LR}^{S}}\right|^2 \Bigg\}.  \eeq
It means that the electron antineutrino beam emitted
in the direction antiparallel to the muon polarization direction
includes only the exotic right-chirality antineutrinos with
$\mbox{\boldmath $\hat{\eta}_{\overline{\nu}}$}\cdot\hat{\bf q}=-1$. 
If the exotic interactions $g_{LR}^S, g_{LR}^T$ are present in the DPMaR, the right-chirality antineutrinos (with negative helicity for $ m_{\overline{\nu}_e} \rightarrow 0$) are no longer "sterile". 

After the integration of the Eqs. (\ref{livetimeV}, \ref{livetimeST},
\ref{livetimeVST}), the  muon  lifetime is as follows:   
\beq
\tau & = & \frac{192 \pi^{3}}{m_{\mu}^5 G_{F}^{2}}\left( \frac{1}{
|g_{LL}^{V}|^2 + \frac{1}{4}|g_{LR}^{S}|^2 + 3|g_{LR}^{T}|^2
}\right). \eeq Because the muon lifetime is measured observable, so
the admittance of the exotic $g_{LR}^{S, T}$ couplings means that
the standard coupling $g_{LL}^{V}$ should be  decreased in order to
the sum $(|g_{LL}^{V}|^2 + \frac{1}{4}|g_{LR}^{S}|^2 +
3|g_{LR}^{T}|^2 )$ was a constant value.\\
 If the antineutrino beam comes from the unpolarized muon decay,
the energy and angular distribution of the
electron antineutrinos consists only of two parts; standard
$(V)$ and exotic $(S + T)$, i. e. Eqs. (\ref{livetimeV},
\ref{livetimeST}) for $\mbox{\boldmath $\hat{\eta}_{\mu}$} \cdot
{\bf \hat{ q}}= 0$. 
 If one puts $E_{\overline{\nu}} = m_\mu/2$ (i. e. $y=1$) in both parts, the standard $(V)$ part vanishes, while the exotic $(S + T)$ one  survives.

\section{Conclusions}
\label{sec3}
In this paper, we have shown that the
  energy and angular distribution of the
electron-antineutrinos in  the muon rest frame can be sensitive to the
interference  terms between the standard left- and exotic right-chirality
antineutrinos, proportional to the transverse components of the
antineutrino beam polarization. The magnitude of the azimuthal asymmetry
caused by the interferences is illustrated in the Fig. 2. The observation
of the  dependence on the angle $\phi$ would be a direct signature of the
right-chirality antineutrinos in the DPMaR.\\
   The admittance of the exotic scalar
and tensor  charged weak interactions  in addition to the standard
vector interaction  in the DPMaR  indicates the possibility of producing
the mixture of the
left- and right-chirality electron antineutrinos with the assigned
direction of the transverse antineutrino spin polarization with
respect to the production plane. Such  polarized beam could be scattered
on
the PET in order to
  measure the CP-violating effects caused by the
  exotic couplings of the right-chirality antineutrinos $g_{LR}^{S, T}$ (or
the
  right-chirality neutrinos) in the purely leptonic processes.
\\We have noticed  that for  $\mbox{\boldmath
$\hat{\eta}_{\mu}$}\cdot\hat{\bf q}=-1$, the energy and angular
distribution of the
electron antineutrinos consists only of the exotic part $(S + T)$. It
means that if the SM prediction is correct, no signal should be detected
for the  electron antineutrino beam emitted
in the direction antiparallel to the muon polarization direction. We see
that the magnitude of this contribution is very small compared with the
dominant one from the standard vector interaction, basing on the current
limits obtained for the non-standard couplings. \\
The DPMaR may also be used to produce the strong left-chirality and
longitudinally polarized (anti)neutrino beam and to measure the dependence
of the antineutrino energy spectrum on the
$\mbox{\boldmath$\hat{\eta}_{\mu}$}\cdot\hat{\bf q}$. So far no such tests
have been carried out.\\
 The observation of the right-handed current interaction is important for
interpreting results on the neutrinoless double beta decay \cite{ndbd}.\\
  We plan to search for the other polarized (anti)neutrino beams, which
could be interesting from the point of observable effects caused by the
exotic right-chirality states.
We expect  some interest in the neutrino laboratories working with
polarized muon decay and
neutrino beams, e.g. KARMEN, PSI, TRIUMF.
\\
\\
{\bf Acknowledgments}
\\
\\
This work was supported in part by the grants of the Polish Commit
tee for Scientific Research LNGS/103/2006 and 1 P03D 005 28.

 \section{ Four-vector antineutrino polarization and density operator} 
 \label{app1}

 The formula for the 
the  spin polarization 4-vector of  massive antineutrino $S^\prime$
moving  with the momentum ${\bf q}$ is as follows:
\beq
S^\prime & = & (S^{\prime 0}, {\bf S^\prime}),\\
S^{\prime 0} & = & \frac{{|\bf q|}}{m_{\overline{\nu}}}(\mbox{\boldmath
$\hat{\eta}_{\overline{\nu}}$}\cdot{\bf \hat{q}}), \\
{\bf S^\prime} & = & -\left(\frac{E_{\overline{\nu}}}{m_{\overline{\nu}}}(\mbox{\boldmath
$\hat{\eta}_{\overline{\nu}}$}\cdot{\bf \hat{q}}){\bf \hat{q}} +
\mbox{\boldmath $\hat{\eta}_{\overline{\nu}}$} - (\mbox{\boldmath
$\hat{\eta}_{\overline{\nu}}$}\cdot{\bf \hat{q}}){\bf \hat{q}}\right),
 \eeq
 where $\mbox{\boldmath $\hat{\eta}_{\overline{\nu}}$}$ - the unit 3-vector of the
  antineutrino polarization in its rest frame.
 The formula for the density operator of the polarized antineutrino
 in the limit of
vanishing  antineutrino mass $m_{\overline{\nu}} $
 is given by:
 \beq
\lim_{m_{\overline{\nu}}\rightarrow 0}\Lambda_{\overline{\nu}}^{(s)} &=&
 \lim_{m_{\overline{\nu}}\rightarrow 0}
\frac{1}{2}\bigg\{\left[(q^{\mu}\gamma_{\mu}) -
m_{\nu}\right]\left[1 +
\gamma_{5}(S^{\prime \mu}\gamma_{\mu})\right]\bigg\}  \\
 & = & \mbox{} \frac{1}{2}\bigg\{(q^{\mu}\gamma_{\mu})
 \left[1 - \gamma_{5}(\mbox{\boldmath
$\hat{\eta}_{\overline{\nu}}$}\cdot{\bf \hat{q}}) + \gamma_{5} (\mbox{\boldmath
$\hat{\eta}_{\overline{\nu}}$} - (\mbox{\boldmath $\hat{\eta}_{\overline{\nu}}$}\cdot{\bf
\hat{q}}){\bf \hat{q}})\cdot
\mbox{\boldmath $\gamma$}\right]\bigg\} \\
 & = & \mbox{} \frac{1}{2}\bigg\{(q^{\mu}\gamma_{\mu})
 \left[1 - \gamma_{5}(\mbox{\boldmath
$\hat{\eta}_{\overline{\nu}}$}\cdot{\bf \hat{q}}) - \gamma_{5}
S^{\prime\perp}\cdot \gamma \right]\bigg\},
\eeq
where $S^{\prime\perp} = \left(0, \mbox{\boldmath $\eta_{\overline{\nu}}
^{\perp}$} =  \mbox{\boldmath $\hat{\eta}_{\overline{\nu}}$} - (\mbox{\boldmath
$\hat{\eta}_{\overline{\nu}}$}\cdot{\bf \hat{q}}){\bf \hat{q}}\right)$.
 We
see that in spite of the singularities $m_{\nu}^{-1}$ in the
polarization four-vector $S^\prime $, the density operator
$\Lambda_{\overline{\nu}}^{(s)}$ remains finite including the transverse
component of the antineutrino spin polarization \cite{Michel}.

\end{document}